\documentclass{llncs}

\usepackage[utf8]{inputenc}
\usepackage{amsmath}
\usepackage{amsfonts}
\usepackage[inference]{semantic} 
\usepackage{latexsym} 
\usepackage{stmaryrd} 
\usepackage{xcolor}
\usepackage{oz} 
\usepackage{wrapfig} 
\usepackage{breakcites} 
\usepackage{xspace}
\usepackage{tikz}
\usetikzlibrary{arrows,backgrounds,automata,shapes,positioning,chains,calc}
\usepackage{appendix}
\usepackage{array} 
\usepackage{hyperref}

\newcommand{\tid}{t}

\newcommand{\TS}{{\it TS}\xspace}

\newcommand{\rdval}{{\it rdval}}
\newcommand{\wrval}{{\it wrval}}

\newcommand{\var}{\mathit{var}}
\newcommand{\vmax}{\mathit{vmax}}

\newcommand{\interf}{interf}
\newcommand{\vps}{\beta}
\newcommand{\sync}{{\it sync}\xspace}
\newcommand{\fence}{fence}

\newcommand{\purple}[1]{{\color{purple}#1}}
\newcommand{\teal}[1]{{\color{teal}#1}}

\newcommand{\Rd}{\mathit{Rd}}

\newcommand{\wlp}{\operatorname{\mathsf{wlp}}}
\newcommand{\dis}{\operatorname{\mathsf{dis}}}

\newcommand{\Act}{{\sf Act}}
\newcommand{\Val}{{\sf Val}}

\newcommand{\Tid}{\mathsf{Tid}}
\newcommand{\VarG}{{\sf Var_G}}
\newcommand{\VarL}{{\sf Var_L}}
\newcommand{\Var}{{\sf Var}}

\newcommand{\as}{\mathcal{A}}
\newcommand{\gl}{\mathcal{G}}
\newcommand{\lo}{\mathcal{L}}

\newcommand{\e}{E}

\newcommand{\reg}{reg}

\definecolor{mycolor}{rgb}{0.122, 0.435, 0.698}

\newcommand{\linefill}{\cleaders\hbox{$\smash{\mkern-2mu\mathord-\mkern-2mu}$}\hfill\vphantom{\lower1pt\hbox{$\rightarrow$}}}  
  
\newcommand{\transi}[2]{\mathrel{\lower1pt\hbox{$\mathrel-_{\vphantom{#2}}\mkern-8mu\stackrel{#1}{\linefill_{\vphantom{#2}}}\mkern-11mu\rightarrow_{#2}$}}}

\newcommand{\Acte}{{\sf Act_{ext}}}

\begin{document}

\title{Lifting the Reasoning Level in Generic Weak Memory Verification (Extended Version)\thanks{Bargmann and Wehrheim are supported by DFG-WE2290/14-1.}}

\author{Lara Bargmann \and
Heike Wehrheim}

\institute{Department of Computing Science, University of Oldenburg, Germany}

\maketitle
\begin{abstract}
Weak memory models specify the semantics of concurrent programs on multi-core architectures. Reasoning techniques for weak memory models are often specialized 
to one fixed model and verification results are hence not transferable to other memory models. 
A recent proposal of a {\em generic} verification technique  based on {\em axioms} on program behaviour expressed 
via weakest preconditions aims at overcoming this specialization to dedicated models. 
Due to the usage of weakest preconditions, reasoning however takes place on a very low level requiring the application of numerous axioms  for deriving program properties, even for a single statement.  

In this paper, we lift reasoning in this generic verification approach to a more  abstract 
level. Based on a view-based assertion language, we provide a number of novel {\em proof rules} for directly reasoning on the level of program constructs. We prove soundness of our proof rules and exemplify them on the write-to-read causality (WRC) litmus test. 
A comparison to the axiom-based low-level proof reveals a significant reduction in the number of required proof steps.

\end{abstract}
\section{Introduction}
\label{sec:int}

The behaviour of concurrent programs running on modern multi-core processors is influenced by the (weak) {\em memory model} of the processor. A memory model fixes how concurrent threads can access shared variables, in particular which values of shared variables a thread can read. The behaviour of weak memory models differs from the often assumed {\em sequential consistency} (SC)~\cite{DBLP:journals/tc/Lamport79} in which an execution is simply an interleaving of sequential executions of threads following their program order. 

As weak memory models deviate from sequential consistency, verification techniques for concurrent programs like rely-guarantee~\cite{DBLP:journals/fac/XuRH97} or Owicki-Gries reasoning~\cite{DBLP:journals/acta/OwickiG76} become unsound on weak memory models. Consequently, past years have seen the development of numerous reasoning approaches {\em specific} to a memory model~(like, e.g.,~\cite{DBLP:conf/esop/BilaDLRW22,ecoop20,DBLP:journals/corr/abs-2004-02983,DBLP:conf/icalp/LahavV15,DBLP:conf/vstte/Ridge10}).  The drawback of all these techniques is that a correctness proof for a concurrent program running on one memory model is not directly transferable to other memory models. 

To alleviate this problem, Doherty et al.~\cite{TOCL} propose a {\em generic} reasoning technique for weak memory models provided these have a {\em view-based semantics}~\cite{DBLP:conf/ppopp/DohertyDWD19}. A view of a thread specifies which write events to shared variables a thread can observe (and hence read from). The core of the reasoning technique is the concept of threads being {\em view-maximal} and memory-model internal steps to not invalidate view-maximality. 
On top of such novel concepts, \cite{TOCL} simply builds on standard Owicki-Gries reasoning for concurrent programs~\cite{DBLP:journals/acta/OwickiG76}. So far, memory models SC, TSO~\cite{DBLP:conf/popl/SarkarSNORBMA09}, PSO~\cite{DBLP:journals/computer/AdveG96} and C11 RAR~\cite{ecoop20} have been shown to fall into this category. Reasoning (about single program instructions) then proceeds by applying low-level axioms based on weakest preconditions. The result is a correctness proof of a concurrent program (a proof outline) which is sound for {\em every} memory model satisfying the axioms. 

While providing a memory-model independent approach, the technique however suffers from the need to apply very low-level, detailed axioms combined with standard properties of weakest preconditions. Moreover, reasoning engines (like Isabelle, as used in~\cite{DBLP:journals/corr/abs-2004-02983}) might not record the axioms employed for a specific proof. 
This hinders transferability to memory models fulfilling only a {\em subset} of the axioms: we do not know anymore whether a proof is or is not valid on such a partially fitting model. 

To improve on these shortcomings, we propose a lifting of the reasoning technique to a higher level. Starting from a view-based language for formulating assertions on concurrent programs, we develop several novel proof rules for program statements. We prove soundness of each of these rules via the low-level axioms. Moreover, together with every new rule we list the required axioms. This enables us to directly see whether a proof is transferable to a memory model which only partially fulfills the axiom set.
We exemplify our new proof rules on the write-to-read causality litmus test (see, e.g.,~\cite{DBLP:conf/pldi/BoehmA08}) for which we  provide both the low-level and the novel high-level reasoning steps. This demonstrates a significant reduction in the number of required proof steps.

\section{Program Syntax}
\label{sec:axiom}

\begin{figure}[t]
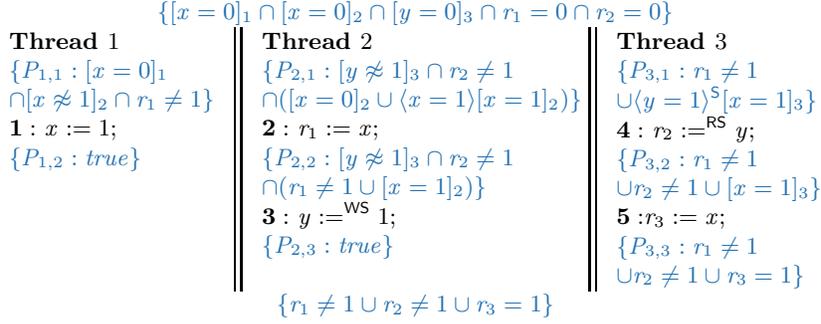

    \begin{center} 
\begin{tabular}[b]{@{}c@{}c@{}} 
   ${\color{mycolor} \{[x=0]_1  \cap [x=0]_2 \cap [y=0]_3 \cap r_1=0 \cap r_2=0\} }$ \\ 
  $\begin{array}{l@{\ \ }||@{\ \ }l||@{\ \ }l}
    \begin{array}[t]{l}
     \textbf{Thread } 1 
     \\
     {\color{mycolor}\{ P_{1,1}: [x=0]_1}\\
     {\color{mycolor}\cap [x\not\approx 1]_2 \cap r_1\neq 1 \} } \\ 
      {\bf 1:} \ x:=1 ;
     \\
    {\color{mycolor}\{ P_{1,2}: true\}  }
     \end{array}
    & 
    \begin{array}[t]{l}
     \textbf{Thread } 2 
     \\
    {\color{mycolor}\{ P_{2,1}: [y\not\approx 1]_3 \cap r_2\neq 1}\\ 
    {\color{mycolor}\cap ([x=0]_2 \cup \langle x=1 \rangle[x=1]_2)  \} } \\
      {\bf 2:}\  r_1:=x ;
     \\
     {\color{mycolor}\{ P_{2,2}: [y\not\approx 1]_3  \cap r_2\neq 1}\\
     {\color{mycolor}\cap (r_1\neq 1 \cup [x=1]_2) \}  }\\
      {\bf 3:}\ y:=^{\sf WS}1 ;
     \\
     {\color{mycolor}\{ P_{2,3}: true\} } 
     \end{array}
     & 
    \begin{array}[t]{l}
     \textbf{Thread } 3
     \\
    {\color{mycolor}\{ P_{3,1}:  r_1\neq 1 }\\
    {\color{mycolor}\cup\langle y = 1\rangle^{\sf S} [x = 1]_3\} } \\
      {\bf 4:}\ r_2:=^{\sf RS}y ;
     \\
     {\color{mycolor}\{ P_{3,2}: r_1\neq 1 }\\
     {\color{mycolor}\cup r_2\neq 1 \cup [x = 1]_3 \} }\\
     {\bf 5:} r_3:=x ;
     \\
     {\color{mycolor}\{ P_{3,3}: r_1\neq 1}\\
     {\color{mycolor}\cup r_2\neq 1 \cup r_3=1 \} } 
     \end{array}
   \end{array}$ 
   \\
   ${\color{mycolor} \{r_1\neq 1  \cup r_2\neq 1 \cup r_3=1\} }$ \\
   \end{tabular} 
\end{center}
\vspace*{-0.5cm}
    \caption{Write-Read-Causality litmus test as proof outline}
    \vspace*{-0.3cm}
    \label{fig:wrc}
\end{figure}
We start by introducing the syntax of concurrent programs.   
We define a concurrent program as a parallel composition of sequential programs. Each thread $t \in \Tid$ runs a sequential program $Com$ and with the function $\Pi:\Tid \to Com$ we model a concurrent program over threads $\Tid$. We let $\VarG$ be the set of global variables and $\VarL$ the set of local variables (or {\em registers}) with $\VarG \cap \VarL=\emptyset$ and $\Var=\VarG\cup \VarL$. We assume that initially all variables have the value 0.

For $x\in \VarG$, $r\in \VarL$ and value $v\in \Val$ the  following grammar defines $Com$: 
\vspace{-1.5mm}
\begin{align*}
E 	&::= v \mid e\\
com & ::= skip \mid fnc \mid r:=E \mid r:=x \mid r:=^{\sf RS} x\mid x:=E \mid x:=^{\sf WS} E\\
Com &::= com \mid Com;Com \mid \text{if } b \text{ then } Com \text{ else } Com \mid \text{while } b \text{ do } Com
\end{align*}

\vspace{-1.5mm}
\noindent where $e\in Exp$ and $b\in BExp$ are expressions over local variables only, $e$ arithmetic and $b$ boolean, and $fnc$ is a so-called {\em fence} (or barrier) instruction.  
The two annotations {\sf RS} (read synchronized) and {\sf WS} (write synchronized) provide ways of synchronizing a write with a read. 

\begin{example}
    In C11~\cite{DBLP:conf/popl/LahavGV16}, the annotation {\sf Release} is  a {\sf WS} 
    and {\sf Acquire} an {\sf RS} synchronization. In SC, reads and writes only occur in synchronized form 
    (and thus the writes and reads without annotation behave like the ones with annotation). Such synchronizations guarantee causal consistency. 
\end{example}

We let $\Act=\{rd(x,r,v),rd^{\sf RS}(x,r,v), wr(x,v), wr^{\sf WS}(x,v), \fence \mid x\in \VarG,$ $r\in \VarL,  v\in \Val \}$ be the set of actions containing synchronized  and unsynchronized reads and writes;  
 $\Acte=\Act \cup \{r:=v \mid r\in Var_L, v\in \Val\} \cup \{\tau\}$ ($\tau$  an internal action). For an action $a\in\Act$,  $\var(a)\in \VarG$ describes the global variable of the action. For a  (synchronized or unsynchronized) read action $rd(x,r,v)$, we let $\rdval(a) = v \in \Val$, otherwise we set $\rdval(a)=\bot\notin \Val$; $\wrval(a)$ is similarly defined for write actions. With these functions we can define subsets of $\Act$: $Rd=\{a\in \Act \mid \wrval(a)=\bot, \rdval(a)\neq \bot\}$ and $Wr=\{a\in \Act \mid  \rdval(a)=\bot, \wrval(a)\neq \bot\}$. For a value $v$, we assume $Rd[v]=\{a\in Rd \mid \rdval(a)=v\}$ and $Wr[v]=\{a\in Wr \mid  \wrval(a)=v\}$. We let $\Act_{|x}$ be the set of all actions $a$ with $\var(a)=x$. Hence $Rd_{|x}$ is the set of all reads in $\Act_{|x}$ and $Wr_{|x}$ the set of all writes. 

\begin{example} Figure~\ref{fig:wrc} shows our running example, the Write-to-Read-Causality litmus test WRC (originally formulated by Boehm and Adve~\cite{DBLP:conf/pldi/BoehmA08} and appearing here in the form of \cite{tutorial}). 
The program is a parallel composition of three threads, accessing global variables $x$ and $y$ plus using local registers $r_1, r_2$ and $r_3$. 
In between statements in the program, we see named assertions (in blue), making it a {\em proof outline}. 
The assertion ${\color{mycolor} \{r_1\neq 1  \cup r_2\neq 1 \cup r_3=1\} }$
at the end (the postcondition) states the expected outcome of the program: when both $r_2$ and $r_1$ take the value 1, then $r_3$ should also be 1. Such a behaviour holds for some but not all memory models, e.g.,~TSO, SC and C11 satisfy it, but, e.g.,~the memory model with a promise semantics in~\cite{DBLP:conf/fm/WehrheimBD23} does not. 
\end{example}

The semantics of programs $Com$ depends on the specific memory model a program runs on. In general, such semantics are typically defined in the following way (see, e.g.,~\cite{TOCL}): First, a semantics for the {\em local} part, i.e.,~the registers, is defined. As registers are not shared among threads, every thread directly writes to and reads from its registers. For shared variables, the local semantics simply assumes that any value can be read. 
In a next step, the local semantics is combined with a specific memory model semantics that details which values can actually be read by which threads in some given state. As we develop a generic reasoning approach here, we cannot further detail the semantics (we have no fixed memory model).

\begin{figure}[t] 
   \begin{align*}
     \Sigma = {} &  \wlp(R,\Sigma) \tag{\text{Non-aborting}}
     \\
     R' \subseteq R \wedge P \subseteq P'  \imp {} &  \wlp(R,P) \subseteq \wlp(R',P')
                                            \tag{\text{(Anti)-Monotonicity}}
     \\
     \wlp(R,\wlp(R',P)) = {} &  \wlp(R \fcmp R',P) \tag{\text{Composition}}
     \\
     R[\wlp(R,P)] \subseteq {} &  P \tag{\text{Relation Application}}
     \\
     \wlp(R,P) \cap {} 
    \wlp(R,Q) = {} &   \wlp(R,P\cap Q) \tag{\text{Conjunctivity}}
     \\
     \wlp(R,P) \cup {} 
    \wlp(R,Q) \subseteq {} &   \wlp(R,P\cup Q) \tag{\text{Disjunctivity}}
   \end{align*}
   \caption{Properties of $\wlp$ ($P, Q \subseteq \Sigma$, $R, R' \subseteq \Sigma \times \Sigma$)}
   \label{fig:wlp} 
\end{figure}

\section{Axiomatic Reasoning} 

Instead of trying to provide separate correctness proofs for WRC for all memory models, we could employ the generic approach in~\cite{TOCL} 
and construct {\em one} proof which is then valid for all memory models fulfilling the axioms employed {\em in this proof}. 
To this end, the generic reasoning technique abstracts from the semantics (and thus from a concrete memory model) and bases reasoning on {\em axioms}. 

\subsection{Axioms}
The approach of~\cite{TOCL} reasons about arbitrary transition systems $TS \sdef (\Acte,\Sigma,$ $I,T)$  where $\Acte$ is the set of actions, $\Sigma$ a set of states, $I \subseteq \Sigma$ a set of initial states and $T \in \Tid \times \Act \rightarrow 2^{\Sigma \times \Sigma}$ a set of transitions. The axiomatisation is build upon the {\em weakest liberal precondition transformer} (wlp)~\cite{DBLP:books/ph/Dijkstra76}, which is used both as a basis for property specification and verification. 
For a relation $R$ and set of states $P$ (representing a predicate), we let 
$\wlp : 2^{\Sigma \times \Sigma} \times 2^\Sigma \rightarrow 2^\Sigma$  
be \vspace{-1.5mm}
$$\wlp(R,P) \sdef \{ \sigma \in \Sigma \mid \forall \sigma' : (\sigma,\sigma') \in R \implies \sigma' \in P \}
$$
Figure~\ref{fig:wlp} details some properties of wlp where $\fcmp$ denotes relational composition and $R[\cdot]$ relational image. Here, $R$ typically
is the relation $T(\tid, a)$, $\tid \in \Tid, a \in \Acte$. 
We say $R$ is {\em disabled} in a state $\sigma$ iff $\sigma \in \dis(R)$ holds, where $\dis(R) \sdef \wlp(R, \emptyset)$. This will in particular be employed for read actions, to state that it is impossible for a thread $\tid$ to read a certain value of a shared variable.

The core concept of reasoning is the idea of {\em views} of threads. In weak memory models, threads {\em observe} global variables to have certain values (namely the values of write actions); a thread might observe several different values at a time and different threads might have different such observations. 
This differs from sequential consistency in which all threads have the same observation and can only see one value at a time. We say that a thread is {\em view maximal}, $\vmax(\tid,a)$ (on an action $a$ operating on a variable $x\in \VarG$), if it has the ``most up-to-date" view on this variable. While non view maximal threads might be able to read older values of $x$, thread $\tid$ reads the most up-to-date value. 

\begin{example}  As an example, consider the WRC program after the execution of line 1 ($x:=1$). In SC, all threads observe $x$ to be 1 (only). In TSO, in which written values are first placed in thread-local store buffers before being flushed to main memory, there is a state in which thread 1 observes $x$ to be 1 while threads 2 and 3 still see $x$ to be 0. 
In such a state, we, e.g.,~have $\dis(T(2, \Rd_{|x}[1])$. 
In C11, there is even a state in which threads 2 and 3 can see $x=1$ and $x=0$ {\em at the  same time}. 
In all these models, we have $\vmax(1,rd(x,\cdot,\cdot))$ (thread 1 is view maximal on $x$) in that state. 
\end{example}

A specific memory model will give rise to some concrete definition of $\vmax$. 
For the axiomatisation it is only important to guarantee that memory model internal steps 
preserve view maximality in the sense of view-preserving simulations.

\begin{definition}\label{def:beta}
For a transition system $\TS=(\Act,\Sigma,I,T)$, a {\em view-preserving simulation},
denoted $\vps$, is the weakest relation $R$ satisfying   for all threads
$\tid \in \Tid$ and all actions $a \in \Act$ \vspace{-1.5mm}
\begin{align}
  \tag{\text{semi-commutation}}
  \label{eqn:vps1}
R \fcmp T (t, a) & \subseteq T (t, a) \fcmp R\\
  \tag{\text{view maximality}}
\label{eqn:vps2}
\vmax(t, a) & \subseteq \wlp(R, \vmax(t, a))
\end{align}

\vspace{-1.5mm}
\end{definition}

\begin{figure}[t]
 \begin{align*}    
     \textbf{C1}: & \forall \tid \in \Tid, a \in \Act: I \subseteq \vmax(\tid, a)  \\
     \textbf{C2}: & \forall \sigma, \sigma' \in \Sigma, \tid, \tid' \in \Tid, a \in \Act: \\ 
    & \sigma \in \vmax(\tid,a) \wedge (\sigma,\sigma') \in T(\tid,a) \imp  \exists
  \tau \in \Sigma:\ (\sigma',\tau) \in \vps \wedge (\sigma,\tau) \in T(\tid',a) \fcmp\vps \\
    \textbf{C3}: & \forall \tid \in \Tid, a\in \Act:  
     T (\tid, a) \subseteq \vps \fcmp \interf (\tid, a) \fcmp \vps \\
  \textbf{C4}: & \forall \tid \in \Tid, a,b \in \Act:  
      \vmax (t, a) \subseteq \wlp(\interf (t, b),\vmax (t, a))
   \end{align*}
    \caption{Core axioms}
    \label{fig:core-axioms}
\end{figure}

A view-preserving simulation keeps view maximality of threads and semi-commutes with the transition relation. 

\begin{example}
    A view-preserving simulation for SC is the identity relation. For TSO it is the flushing of contents of store buffers to main memory. For C11 in which all write events to the same variable $x$ are ordered in some {\em modification order}, it is the advancement of a thread's observation on $x$ (a write to $x$ of a value) to another write which occurs later in modification order. 
\end{example}

The concept of views is inherent to the axiomatic reasoning and hence is also employed for property specification. As threads might observe more than one value for a variable, the ordinary first-order logic assertions on program variables of Hoare-logic~\cite{DBLP:journals/cacm/Hoare69} need to be replaced by {\em view-based} assertions. 

\begin{definition} 
\label{def:assertions}
For a thread $t$, a variable $x \in \VarG$ and values $u,v \in \Val$ we define

\vspace{0.5em}
$\begin{array}{@{}r@{~}l@{\ \ }l} 
    [x \not \approx v]_t &\sdef \dis(T(t, \Rd_{|x}[v])) & \text{(Impossible value)} \\  
    {[}x \equiv v]_t
                         &\sdef \bigcap_{u \neq v} [x \not \approx u]_t  &\text{(Definite value)}\\
    x_{\uparrow t}  & \sdef \bigcap_{a\in \Act_{|x}} \vmax(t,a) 
    & \text{(Maximal view)} \\ 
   {[}x = v]_t 
&\sdef {[}x \equiv v]_t \cap x_{\uparrow t} & \text{(Maximal value)} \\ 
 \langle y = u \rangle^{\sf S} [x = v]_t & \sdef \wlp(T(t, rd^{\sf RS}(y,\cdot,u)), [x=v]_t)  & \text{(Synced conditional}\\ 
  & & \text{ observation)} \\
 \langle x = v \rangle [x = v]_t & \sdef \wlp(T(t,rd(x,\cdot,v)), [x=v]_t) & \text{(Conditional} \\
  & & \text{ observation)}
\end{array}$
\end{definition}

\begin{figure}[t]
 \begin{align*}    
    \textbf{SV1}: &  \forall a,b \in \Act, \tid, \tid'\in \Tid \text{ s.t. } \var(a) \neq \var(b): \\ 
         & \interf (\tid', b) \fcmp T (\tid, a) \subseteq T (\tid, a) \fcmp \interf (\tid', b) \\
    \textbf{SV2}: & \forall a,b \in \Act, t,t'\in \Tid \text{ s.t. } \var(a) \neq \var(b):  \\
         & \vmax (t,a) \subseteq \wlp(\interf (t', b), \vmax (t,a)) \\
    \textbf{RW1}: & \forall t,t' \!\in\! \Tid, x \!\in\! \VarG, a_r\! \in\! \Rd_{|x}, a_w\! \in\! Wr_{{|}x} \text{ s.t. } \rdval(a_r) \not =  \wrval(a_w): \\ 
         & \interf(t', a_w) \fcmp T(t, a_r) \subseteq T(t, a_r) \fcmp \interf(t', a_w) \\
    \textbf{RW2}: & \forall a\in \Act, t,t'\in \Tid, a_r \in \Rd_{|\var(a)}:\\ & \interf(t', a_r) \fcmp T(t,a) \subseteq T(t,a) \fcmp \interf(t', a_r) \\
    \textbf{RW3}: & \forall a \in \Act, t,t'\in \Tid, a_r \in \Rd_{|\var(a)}:\\  & \vmax (t, a) \subseteq \wlp(\interf(t', a_r), \vmax (t, a)) \\
    \textbf{RW4}: & \forall x\in \VarG, t \in \Tid:  \Sigma \subseteq \dom (T (t, Rd_{|x})) \\
    \textbf{RW5}: & \forall x \in \VarG, a_w \in Wr_{|x}, v = \wrval(a_w):  \Sigma \subseteq \wlp(T(t, a_w), \dom (T (t, \Rd_{|x}[v]))) \\
    \textbf{RW6}: & \forall x\in \VarG, t \in \Tid: 
          x_{\uparrow t} \subseteq \bigcup_{v \in \Val} {[}x \equiv v]_t  \\
    \textbf{RW7}: & \forall x \in \VarG, a_w, a_r, a\in \Act_{|x}, t, t' \in \Tid, \text{s.t.}\, \wrval(a_w) = \rdval(a_r) \wedge t \neq t':\\&
          \vmax(t, a_w) \cap \dis(T(t',a_r)) \subseteq \wlp(T(t,a_w), \wlp(T(t', a_r), \vmax(t', a)))
   \end{align*} 
    \caption{Axioms on shared variables} 
    \label{fig:sv-axioms}
\end{figure}

\begin{example}
    Consider the state of WRC after executing lines 1, 2 and 3 (in this order).   In SC, we then have $[x=1]_{\tid}$ for all threads $\tid$ (same for $y$). 
    In $TSO$ (when store buffer contents has not been flushed yet), we, e.g.,~have $[x=1]_1$, $[y=1]_2$ and $[x \not \approx 1]_3$ (thread 3 cannot read $x$ to be 1). In C11, we might have $[y \approx 0]_3$ and $[y \approx 1]_3$ (thread 3 can read both 0 and 1). Moreover, the following synced conditional observation is valid in all three memory models: $\langle y=1 \rangle^{\sf S} [x=1]_3$ (by a synchronized read of $y$ to be 1, thread 3 becomes view maximal on $x$ and definitely observes the value 1 for $x$). 
\end{example}

We let $\gl$ be the set of (all logical combinations of) such {\em global} assertions. In our proof outlines (like in the one of WRC) we also allow for normal Hoare-like assertions on local registers (e.g.~$(r_1 = 1) \in BExp$), and define the logical combinations of such {\em local} ($\lo$) assertions and the global assertions to be the set $\as$ of all assertions.

Assertions define sets of states. Of particular interest are $\vps$-stable assertions. 

\begin{definition}
Any predicate $P \in 2^\Sigma$ is {\em $\vps$-stable} iff
$P \subseteq \wlp(\vps,P)$. 
\end{definition}

All assertions in  $\gl$ are $\vps$-stable (see~\cite{TOCL}). 
The axioms furthermore make use of an {\em interference relation} $\interf \in \Tid \times \Act \to 2^{\Sigma  \times \Sigma}$ which (together with $\vps$) provides an {\em overapproximation} of the transition relation $T(\tid,a)$ in order to abstract from details of the memory model and to regain standard properties of reasoning (like writes and reads on different variables commuting). Figure~\ref{fig:core-axioms} gives all core axioms; Figure~\ref{fig:sv-axioms} gives axioms concerning read and write actions on shared variables. 

We only briefly explain the axioms; an example application of the axioms for reasoning about WRC is given below.  
Axiom {\bf C1} states that initially all threads are view maximal w.r.t.~all actions. 
Axiom {\bf C2} describe the independence of actions w.r.t.~thread identifiers (where additional $\vps$ steps are required). 
Axiom {\bf C3} states that $\interf$ together with $\vps$ 
{\em over-approximates} the behaviour of an action. 
Axiom {\bf C4}  states that the interference relation preserves every
view-maximality property of the thread performing the interference (of the action).

Axiom {\bf SV1} is a weakening of the commutation property present in SC. 
{\bf SV2}  states that a view-maximality property of any thread
is stable under actions on any other variable. 
Axioms {\bf RW1} and {\bf RW2} capture semi-commutativity properties for writes and reads, respectively, and are analogous to {\bf SV1}. 
Axiom {\bf RW3} states that view-maximality on a variable is preserved by reading the variable. 
Axiom {\bf RW4} states that it is always possible to read some value of a variable, and {\bf RW5} states that a thread $t$ 
writing some value can afterwards read it.
{\bf RW6} states that whenever $t$ is view maximal on actions over variable $x$, then $t$ has a definite value assertion over {\em some} value for $x$ (i.e., can only read one value for $x$).  
Axiom {\bf RW7} considers a situation in which thread $t$ is $\vmax$ on a variable $x$ but $t'$ cannot read a specific value for this variable. We then obtain view-maximality of $t'$ on $x$ after $t$ has performed the write $a_w$ and $t'$ has read this write's value.

Finally, the axiom set contains one specific axiom for {\em fences}  and one for {\em message passing}. 
Fence instructions are employed in weak memory models to make programs behave more like SC. 
The fence axiom given below states this by saying that a fence in a thread being view maximal on some action $a$ makes all other threads view maximal on $a$ as well.

\begin{description}
  \item[FNC]
     $\forall a\in \Act,\, t,t' \in \Tid$: $\vmax(t,a) \subseteq \wlp(T(t,\fence),\vmax(t',a))$.
   \item[MP] 
  For $a_w,a_r, b \in \Act$ and $\purple{t},\teal{t'} \in \Tid$ such that ${\color{blue} (a_w,a_r) \in \sync}$, $\var(a_w) = \var(a_r)$,  $\wrval(a_w) = \rdval(a_r)$, $\var(b) \neq \var(a_w)$, and $\purple{t} \neq \teal{t'}$, we have      \vspace{-1.5mm}
  \begin{align*}
      &\vmax(\purple{t}, b) \cap \wlp(\teal{T(\teal{t'}, a_r)}, \vmax (\teal{t'}, b)) \\
      & \quad \subseteq
      \wlp(\purple{T(t, a_w)}, \wlp(\teal{T(t', a_r)}, \vmax (\teal{t'}, b))). 
  \end{align*}
\end{description} 

The message passing axiom \textbf{MP} describes the {\em passing of knowledge on variable values} from one thread to another upon {\em synchronization}. Synchronization is incorporated here by requiring ${\color{blue}(a_w,a_r) \in sync}$ which is achieved when the write has a {\sf WS} and the read an {\sf RS} annotation. 
More specifically, it describes a situation where a  thread $\purple{t}$ is maximal on some action $b$ ($\vmax(\purple{t},b)$) and thread $\teal{t'}$ upon executing action $\teal{a_r}$ would become view maximal on $b$ as well. 
Then, writing the value to be read (i.e., $\purple{T(t, a_w)}$) followed by reading this value ($\teal{T(t', a_r)}$) makes 
thread $\teal{t'}$ view maximal on $b$.

As a first result, we restate two lemmas stating the stability of global assertions under fence and read actions.

\begin{lemma}[\cite{pso-generic}]\label{lem:fencestable}  
    Assume the axioms \textbf{C3}, \textbf{SV1} and \textbf{SV2} hold. For all $P\in\gl$ and threads $t$,   $P\subseteq \wlp(T(t,\fence),P)$.
\end{lemma}

\begin{lemma}[\cite{TOCL}]\label{lem:readstable}
    Assume the axioms \textbf{C3}, \textbf{SV1}, \textbf{SV2}, \textbf{RW2} and \textbf{RW3} hold. For all  $P\in\gl$, threads $t$ and $a_r\in \Rd$,    $P\subseteq \wlp(T(t,a_r),P)$.
\end{lemma}

Note that -- contrary to \cite{pso-generic,TOCL} -- we name the axioms required for the proof in the lemmata.
This is of importance for dealing with memory models which only fulfill part of the axioms (so that we can see whether a generic proof is transferable to such a memory model). 

\subsection{Reasoning Example on Axiom Level}

Next, we employ the axioms for showing one step in the correctness proof of WRC. Note that the proof of WRC in the generic framework has not appeared before. 
In general, such proofs involve proof steps of the form 
\vspace{-1.5mm}
$$P\subseteq\wlp(T(t,a),Q)$$ 

\vspace{-1.5mm}
\noindent for actions $a$ belonging to program instructions $com_t$, where $P \in \as$ is the pre-assertion before and $Q\in \as$ the post-assertion after the instruction.  
We also write these as Hoare-triples 
\vspace{-1.5mm}
$$\{P\}\ com_t\ \{Q\}\ .$$

\vspace{-1.5mm}
\noindent Such steps need to be performed to show local and global correctness (as of Owicki-Gries' approach~\cite{DBLP:journals/acta/OwickiG76}). 

\begin{definition} \label{def:correct} 
 A thread $t$ is {\em locally correct} in a proof outline if $\{P\}com_t\{Q\}$ holds for every program command $com$ in $t$ with pre-assertion $P$ and post-assertion $Q$.  
 
 A proof outline is {\em globally correct} (interference-free) if for every pair of threads $t, t'$,  $\{R\cap P\}com_{t'}\{R\}$ holds for every assertion $R$ in the proof outline of $t$ and command $com$ with pre-assertion $P$ in thread $t'$.
\end{definition}

We exemplify one such proof step for the proof outline in Fig. \ref{fig:wrc}, which is part of the local correctness of thread 3. 
\begin{align*}
    {\color{mycolor} \{r_1\neq 1\cup \langle y=1\rangle^{\sf S}[x=1]_3 \}}\ r_2:=^{\sf RS}_3 y\ {\color{mycolor} \{r_1\neq 1\cup r_2\neq 1\cup [x=1]_3\}}
\end{align*}
For this we have to prove for every $v\in\Val$
$$r_1\neq 1\cup \langle y=1\rangle^{\sf S}[x=1]_3 \subseteq \wlp(T(3,rd^{\sf RS}(y,r_2,v)),r_1\neq 1\cup r_2\neq 1\cup [x=1]_3)$$
Because of the disjunctivity of $\wlp$ (see Fig. \ref{fig:wlp}), we can divide the proof in two parts
\begin{itemize}
    \item[(i)]$r_1\neq 1 \subseteq \wlp(T(3,rd^{\sf RS}(y,r_2,v)),r_1\neq 1)$
    \item[(ii)]$\langle y=1\rangle^{\sf S}[x=1]_3 \subseteq \wlp(T(3,rd^{\sf RS}(y,r_2,v)), r_2\neq 1\cup [x=1]_3)$
\end{itemize}
For reasoning about local registers, we employ a version of the standard technique of backward substitution from the rule of assignment of Hoare-logic\footnote{Such backward substitution is sound here as it only considers local registers.}, i.e., 
$$ e[r:=v] \subseteq \wlp(T(t,rd^{\sf RS}(x,r,v))),e) $$
where $e \in Exp$ is an expression on local variables only and $[r:=v]$ means replacing all occurrence of $r$ by value $v$. 
For (i) we then have  
$$(r_1\neq 1) = (r_1\neq 1[r_2:=v]) \subseteq \wlp(T(3, rd^{\sf RS}(y,r_2,v)),r_1\neq 1)$$
For (ii) we look at two cases. First, let $v=1$. Using the monoticity of $\wlp$ we get
\begin{align*}
    \langle y=1\rangle^{\sf S}[x=1]_3   &=          \wlp(T(3, rd^{\sf RS}(y,r_2,1)), [x=1]_3)\\
                                &\subseteq  \wlp(T(3,rd^{\sf RS}(y,r_2,v)), r_2\neq 1\cup [x=1]_3)
\end{align*}
In the case $v\neq 1$, we need Lemma \ref{lem:readstable} and therefore the axioms C3, SV1, SV2, RW2, and RW3 have to hold. Because of the disjunctivity of $\wlp$ we get 
\begin{align*}
    \langle y=1\rangle^{\sf S}[x=1]_3   &\subseteq  \Sigma\\
                                &\overset{v\neq 1}{=}          (r_2\neq 1[r_2:=v]) \cup [x=1]_3\\
                                &\subseteq  \wlp(T(3,rd^{\sf RS}(y,r_2,v)), r_2\neq 1\cup [x=1]_3)
\end{align*}

Many steps of such correctness proofs are complex, time consuming and repetitive. In the next section we summarize multiple such steps into {\em proof rules} and thereby lift reasoning to the higher level of syntactic assertions, not employing  
 weakest preconditions anymore. 

\section{Rules}
\label{sec:rul}

\begin{figure}[t]
\footnotesize
    $$\inference[\textsc{True}]{  }
      { \{P\}\ com_t\ \{true\} }
    \quad \inference[\textsc{False}]{  }
      { \{false\}\ com_t\ \{P\} }$$
    $$\inference[\textsc{Mono}]{ P_1\supseteq P_2 \quad Q_1\subseteq Q_2 \quad \{P_1\}\ com_t\ \{ Q_1 \} }
      { \{P_2\}\ com_t\ \{ Q_2 \} }$$
    $$\inference[\textsc{Conj}]{ \{P_1\}\ com_t\ \{ Q_1 \} \quad \{P_2\}\ com_t\ \{ Q_2 \} }
      { \{P_1 \cap P_2\}\ com_t\ \{Q_1 \cap Q_2\} }$$
    $$\inference[\textsc{Disj}]{ \{P_1\}\ com_t\ \{ Q_1 \} \quad \{P_2\}\ com_t\ \{ Q_2 \} }
      { \{P_1 \cup P_2\}\ com_t\ \{Q_1 \cup Q_2\} }$$
    \caption{General rules}
    \label{fig:genrules}
\end{figure}

In this section we explain our novel proof rules for the axiomatic reasoning. 
Remember that for a program command $com_t$ in a thread $t$, we prove 
$\{P\}\ com_t\ \{Q\}$
for assertions $P,Q\in\as$ by showing
\begin{align*}
    P\subseteq \wlp(T(t,a), Q)
\end{align*}
where $a$ is the action in $com_t$.
Some interim results of those proofs can be generalised and lifted to the higher level of syntactic assertions. 
We formalise them in the form of rules which then can be used to directly prove the correctness of a proof outline without the need of weakest preconditions. 

We start by giving general rules (Fig.~\ref{fig:genrules}) which hold regardless of the validity of axioms.
Those rules are all in the original Hoare-logic form \cite{DBLP:journals/cacm/Hoare69} and are here translated to our setting. 
For the rules \textsc{True} and \textsc{False} note that the assertions $\{true\}$ and $\{false\}$ describe the set of states $\Sigma$ and the empty set, respectively. 
With that in mind both rules follow directly from our definition of Hoare-triple. 
The intuitive idea of the \textsc{Mono} rule are that a Hoare-triple still holds if the pre-assertion becomes stronger or the post-assertion weaker. 
The first follows by definition and the second from the monotonicity of $\wlp$ (see Fig.~\ref{fig:wlp}). 
Analogously, the rules \textsc{Conj} and \textsc{Disj} formalise the conjunctivity and disjunctivity properties of Figure~\ref{fig:wlp}. 
Hence we get the following Theorem. 
\begin{theorem}\label{th:general}
    The general proof rules in Figure~\ref{fig:genrules} are sound. 
\end{theorem}
The proof of the theorem can be found in the appendix. 
Note that these rules can be used to combine different Hoare-triples from other rules. 

\begin{figure}[t]
\footnotesize
    $$\inference[\textsc{Fence1(Lemma \ref{lem:fencestable})}]{ P\in\gl \quad C3 \quad SV1 \quad SV2 }
      { \{P\}\ fnc_t\ \{P\} }
    \quad \inference[\textsc{Fence2}]{ FNC }
      { \{x_{\uparrow t}\}\ fnc_t\ \{x_{\uparrow t'}\} }$$
    $$\inference[\textsc{Fence3}]{ C2 \quad C3 \quad SV1 \quad SV2 \quad RW6 \quad FNC }
      { \{[x=v]_t\}\ fnc_t\ \{[x=v]_{t'}\} }$$ 
    \caption{Fence rules}
    \label{fig:fencerules}
\end{figure}

Next we look at rules specific to a certain program command and start with fence actions. 
If we formalise the property given in Lemma \ref{lem:fencestable}, we get the first rule of Figure \ref{fig:fencerules}: \textsc{Fence1}. 
Note that with regard to showing global correctness, the rule implies the following lemma.
\begin{lemma}
    \label{lem:globalfence}
    In every proof outline fence actions are globally correct for $\beta$-stable assertions, i.e., for every assertions $G\in \gl$ and $P\in \as$: $\{G\cap P\}\ fnc_t\ \{G\}$. 
\end{lemma}
The \textbf{FNC} Axiom is formalised in \textsc{Fence2} and if we additionally assume the axioms \textbf{C2} and \textbf{RW6} we can not only pass view-maximality to a different thread, but also the value that can be read. 
In the appendix we show the following theorem.
\begin{theorem}\label{th:fence}
    The fence proof rules in Figure~\ref{fig:fencerules} are sound.
\end{theorem}

\begin{figure}[t]
\footnotesize
    $$\inference[\textsc{Read1(Lemma \ref{lem:readstable})}]{ P\in \gl \quad C3 \quad SV1 \quad SV2 \quad RW2 \quad RW3 }
      { \{P\}\ r:=_t x\ \{P\} }$$
    $$\inference[\sc{Read2}]{ }{\{[x\not\approx v]_t\}\ r:=_t x\ \{r\neq v\}}
    \quad\inference[\textsc{Read3}]{  }
      { \{[x=v]_t\}\ r:=_t x\ \{r=v\} }$$
      $$\inference[\textsc{ConRead1}]{  }
      { \{\langle x=v \rangle [x=v]_t\}\ r:=_t x\ \{r\neq v \cup [x=v]_t\} } $$
    $$\inference[\textsc{ConRead2}]{  }
      { \{\langle x=v \rangle^{\sf S} [y=u]_t\}\ r:=^{\sf{RS}}_t x\ \{r\neq v \cup [y=u]_t\} } $$
    $$\inference[\textsc{ReadReg}]{ P\in\lo \quad r\not\in\reg(P)  }
      { \{P\}\ r:=_t x\ \{P\} }
    \quad\inference[\sc{LocRead}]{r\not\in \reg(P)}
      {\{P\}\ r:=_t \e\ \{P\}}$$
    \caption{Read rules ($reg(P)$ being the local registers occurring in $P$) }
    \label{fig:readrules}
\end{figure}

For read actions (Fig.\ref{fig:readrules}) we similarly formalise Lemma \ref{lem:readstable} in rule \textsc{Read1} and get the following lemma.
\begin{lemma}
    \label{lem:globalread}
    In every proof outline read actions are globally correct for $\beta$-stable assertions, i.e., for every assertions $G\in \gl$ and $P\in \as$: $\{G\cap P\}\ r:=_t v\ \{G\}$.
\end{lemma}
The rules \textsc{Read2, Read3, ConRead1} and \textsc{ConRead2} describe how we replace different global assertions (containing $x$) by local ones (containing $r$) after reading the value of $x$ to $r$. 
Here \textsc{Read2} says that if thread $t$ cannot read $v$ for $x$, then after reading $x$ to $r$, $r$ cannot be $v$. 
Analogous in \textsc{Read3} where $t$ cannot read a value different from $v$ for $x$ and is view maximal (which means that $t$ can read the most up-to-date value for $x$), after the read, $r$ has to be equal to $v$. 
If we have a conditional observation assertion $\langle x=v\rangle[x=v]_t$ 
and read in the same thread from $x$, then either we do not read $v$ or $[x=v]_t$ holds afterwards (\textsc{ConRead1}).
We get a similar rule for the synchronized read and the synced conditional observation (\textsc{ConRead2}).
\textsc{ReadReg} tells us that a local assertion remains unchanged after a read to a register which is not included in the assertion.
In \textsc{LocRead} we describe that an assertion will not change, if we read a local expression to a register. In this case the register must not be included in the assertion. 
Note that by $reg(P)$ we mean the set of registers in $P$.
Summarised we get
\begin{theorem}\label{th:read}
    The read proof rules in Figure~\ref{fig:readrules} are sound.
\end{theorem}
which we also proved in the appendix. 

\begin{figure}[t] 
\footnotesize 
	$$\inference[\textsc{Write1}]{ x \neq y \quad C3 \quad SV1 }
      { \{[y\not \approx u]_{t'}\}\ x:=_t E\ \{[y\not \approx u]_{t'}\} }$$
    $$\inference[\textsc{Write2}]{ x \neq y \quad C3 \quad SV1 }
      { \{[y\equiv u]_{t'}\}\ x:=_t E\ \{[y\equiv u]_{t'}\} }$$
    $$\inference[\textsc{Write3}]{ x \neq y \quad C3 \quad SV2 }
      { \{y_{\uparrow t'}\}\ x:=_t E\ \{y_{\uparrow t'}\} }
    \quad \inference[\textsc{Write4}]{ x \neq y \quad C3 \quad SV1 \quad SV2 }
      { \{[y=u]_{t'}\}\ x:=_t E\ \{[y=u]_{t'}\} }$$
    $$\inference[\textsc{Write5}]{ C3 \quad C4 }
      { \{x_{\uparrow t}\}\ x:=_t E\ \{x_{\uparrow t}\} }
    \quad \inference[\textsc{Write6}]{\llbracket E \rrbracket=v \quad C3 \quad C4 \quad RW5 \quad RW6 }
      { \{[x = u]_t\}\ x:=_t E\ \{[x = v]_t\} }$$
    $$\inference[\textsc{ConWrite1}]{\llbracket \e \rrbracket=v \quad t\neq t' \quad C2 \quad C3 \quad C4\\ SV1 \quad SV2 \quad RW2 \quad RW3 \quad RW5 \quad RW6 \quad RW7 }
      { \{[x\not \approx v]_{t'} \cap [x=u]_{t}\}\ x:=_t \e\ \{\langle x=v \rangle [x=v]_{t'}\} } $$
    $$\inference[\textsc{ConWrite2}]{\llbracket \e \rrbracket=v \quad t\neq t' \quad x\neq y \quad C2 \quad C3\\ SV1 \quad SV2 \quad RW2 \quad RW3 \quad RW6 \quad MP }
      { \{[x\not \approx v]_{t'} \cap [y=u]_{t}\}\ x:=^{\sf{WS}}_t \e\ \{\langle x=v \rangle^{\sf S} [y=u]_{t'}\} } $$
    $$\inference[\textsc{WriteReg}]{ P\in\lo }
      { \{P\}\ x:=_t E\ \{P\} } $$
\caption{Write rules}
 \label{fig:writerules} 
\end{figure}

In Figure \ref{fig:writerules} we formalised rules for write actions. 
There we differentiate between  global assertions about the variable written to and about other variables. 
In both cases we need the \textbf{C3} axiom.
This allows us to apply a hand full of axioms that describe properties of $\interf$, e.g., \textbf{SV1} and \textbf{SV2}.
For different variables we can pass readability of a value with the axiom \textbf{SV1} (see rules \textsc{Write1} and \textsc{Write2}). 
If we want to pass view-maximality (\textsc{Write3}), we need \textbf{SV2}.
The rule \textsc{Write4} combines \textsc{Write2} and \textsc{Write3}.
In the case where the assertion contains the same variable as the write action, we can use the axiom \textbf{C4} to pass view-maximality (\textsc{Write5}).
If we additionally assume \textbf{RW5} and \textbf{RW6}, we can update the value thread $t$ can read (\textsc{Write6}). 
If we write a new value to $x$ (which means that before the write, $t$ and $t'$ could not read $v$) in a view-maximal thread $t$, then if $t'$ can read $v$, it also has to be view-maximal. This behaviour is decribed in rule \textsc{ConWrite1}. We need to assume \textbf{RW7} to pass the conditional view-maximality to a different thread. 
The rule \textsc{ConWrite2} describes message passing. 
If $t'$ can read $v$ for $x$ and $t$ can read $u$ for $y$ and is view-maximal in $t$, then if we write $v$ to $x$ in $t$, $[y=u]_{t'}$ only holds if we can read $v$ for $x$ in $t'$. 
This behaviour only differs from \textsc{ConWrite1} by allowing different variables. Because of this, we cannot apply \textbf{RW7} and need \textbf{MP}. Hence this rule only holds for synchronised writes.
The last rule of Figure \ref{fig:writerules} (\textsc{WriteReg}) formalises the fact that a write will not change the value of a register. 
In the appendix, we prove 
\begin{theorem}\label{th:write}
    The write proof rules in Figure~\ref{fig:writerules} are sound.
\end{theorem}
With all these rules being sound, we can now prove correctness much easier and shorter. Also we then know exactly which axioms we need for a certain proof outline to be valid. 

\section{Correctness Proof of WRC via Proof Rules}
\label{sec:exp}

\begin{table}[t]
\caption{Rules employed for showing local correctness of the WRC proof outline}   
    \label{tab:local}
\centering
    \begin{tabular}{ | l | c | c | c | } 
    \hline
    $com_t$ & Hoare-Triples & Proof Rules & Axioms \\ 
    \hline
    $x:=_1 1$ & $\{P_{1,1}\}\ x:=_1 1\ \{P_{1,2}\}$ & \sc{True} & \\
    \hline
    $r_1:=_2 x$ & $\{P_{2,1}\}\ r_1:=_2x\ \{P_{2,2}\}$ & \sc{Read1, ReadReg, } & C3, SV1, SV2, \\
     & & \sc{Read3, ConRead1} & RW2, RW3\\
    \hline
    $y:=_2^{\sf WS}1$ & $\{P_{2,2}\}\ y:=_2 1\ \{P_{2,3}\}$ & \sc{True} &\\
    \hline
    $r_2:=_3^{\sf RS}y$ & $\{P_{3,1}\}\ r_2:=_3 y\ \{P_{3,2}\}$ & \sc{ReadReg, ConRead2} & \\
    \hline
    $r_3:=_3 x$ & $\{P_{3,2}\}\ r_3:=_3 x\ \{P_{3,3}\}$ & \sc{ReadReg, Read3} & \\
    \hline
    \end{tabular}
    
\end{table}

\begin{table}
\caption{Rules employed for showing global correctness of the WRC proof outline}   
    \label{tab:global}
\centering
    \begin{tabular}{ | l | c | c | c | } 
    \hline
    $com_t$ & Hoare-Triples & Proof Rules & Axioms \\ 
    \hline
    $x:=_1 1$ & $\{P_{1,1} \cap P_{2,1}\}\ x:=_1 1\ \{P_{2,1}\}$ & \sc{Write1,} & C2, C3, C4, SV1,\\
     &  & \sc{WriteReg,} & SV2, RW2, RW3,\\
     &  & \sc{ConWrite1} & RW5, RW6, RW7 \\
     & $\{P_{1,1} \cap P_{2,2}\}\ x:=_1 1\ \{P_{2,2}\}$ & \sc{Write1, WriteReg} & C3, SV1\\
     & $\{P_{1,1} \cap P_{2,3}\}\ x:=_1 1\ \{P_{2,3}\}$ & \sc{True} & \\
     & $\{P_{1,1} \cap P_{3,1}\}\ x:=_1 1\ \{P_{3,1}\}$ & \sc{WriteReg} & \\
     & $\{P_{1,1} \cap P_{3,2}\}\ x:=_1 1\ \{P_{3,2}\}$ & \sc{WriteReg} & \\
     & $\{P_{1,1} \cap P_{3,3}\}\ x:=_1 1\ \{P_{3,3}\}$ & \sc{WriteReg} & \\
    \hline
    $r_1:=_2 x$ & $\{P_{2,1} \cap P_{1,1}\}\ r_1:=_2 x\ \{P_{1,1}\}$ & \sc{Read1,} & C3, SV1, SV2,\\
     & & \sc{Read3} & RW2, RW3 \\
     & $\{P_{2,1} \cap P_{1,2}\}\ r_1:=_2 x\ \{P_{1,2}\}$ & \sc{True} & \\
     & $\{P_{2,1} \cap P_{3,1}\}\ r_1:=_2 x\ \{P_{3,1}\}$ & \sc{Read1} & C3, SV1, SV2,\\
     & & & RW2, RW3 \\
     & $\{P_{2,1} \cap P_{3,2}\}\ r_1:=_2 x\ \{P_{3,2}\}$ & \sc{ReadReg} &  \\
     & $\{P_{2,1} \cap P_{3,3}\}\ r_1:=_2 x\ \{P_{3,3}\}$ & \sc{ReadReg} &  \\
    \hline
    $y:=_2^{\sf SW}1$ & $\{P_{2,2} \cap P_{1,1}\}\ y:=_2 1\ \{P_{1,1}\}$ & \sc{Write1, Write4, } & C3, SV1, SV2\\
     &  & \sc{WriteReg} & \\
     & $\{P_{2,2} \cap P_{1,2}\}\ y:=_2 1\ \{P_{1,2}\}$ & \sc{True} & \\
     & $\{P_{2,2} \cap P_{3,1}\}\ y:=_2 1\ \{P_{3,1}\}$ & \sc{WriteReg,} & C2, C3, SV1, SV2, \\
     & & \sc{ConWrite2} & RW2, RW3, RW6, MP\\
     & $\{P_{2,2} \cap P_{3,2}\}\ y:=_2 1\ \{P_{3,2}\}$ & \sc{WriteReg} & \\
     & $\{P_{2,2} \cap P_{3,3}\}\ y:=_2 1\ \{P_{3,3}\}$ & \sc{WriteReg} & \\
    \hline
    $r_2:=_3^{\sf SR} y$ & $\{P_{3,1} \cap P_{1,1}\}\ r_2:=_3 y\ \{P_{1,1}\}$ & \sc{Read1, } & C3, SV1, SV2,\\
     & & \sc{ReadReg} & RW2, RW3 \\
     & $\{P_{3,1} \cap P_{1,2}\}\ r_2:=_3 y\ \{P_{1,2}\}$ & \sc{True} & \\
     & $\{P_{3,1} \cap P_{2,1}\}\ r_2:=_3 y\ \{P_{2,1}\}$ & \sc{Read1, } & C3, SV1, SV2, \\
     & & \sc{Read2} & RW2, RW3\\
     & $\{P_{3,1} \cap P_{2,2}\}\ r_2:=_3 y\ \{P_{2,2}\}$ & \sc{Read1, ReadReg,} & C3, SV1, SV2, \\
     &  & \sc{Read2} & RW2, RW3\\
     & $\{P_{3,1} \cap P_{2,3}\}\ r_2:=_3 y\ \{P_{2,3}\}$ & \sc{True} & \\
    \hline
    $r_3:=_3x$ & $\{P_{3,1} \cap P_{1,1}\}\ r_3:=_3 x\ \{P_{1,1}\}$ & \sc{Read1, } & C3, SV1, SV2, \\
     & & \sc{ReadReg} & RW2, RW3\\
     & $\{P_{3,1} \cap P_{1,2}\}\ r_3:=_3 x\ \{P_{1,2}\}$ & \sc{True} & \\
     & $\{P_{3,1} \cap P_{2,1}\}\ r_3:=_3 x\ \{P_{2,1}\}$ & \sc{Read1,} & C3, SV1, SV2, \\
     & & \sc{ReadReg} & RW2, RW3\\
     & $\{P_{3,1} \cap P_{2,2}\}\ r_3:=_3 x\ \{P_{2,2}\}$ & \sc{Read1,} & C3, SV1, SV2, \\
     & & \sc{ReadReg} & RW2, RW3\\
     & $\{P_{3,1} \cap P_{2,3}\}\ r_3:=_3 x\ \{P_{2,3}\}$ & \sc{True} & \\
    \hline
    \end{tabular}
    
\end{table}

In this section, we finally apply our rules to the correctness proof of the WRC example in Figure~\ref{fig:wrc}. 

\begin{lemma}\label{th:proofoutline}
The proof outline in Figure~\ref{fig:wrc} is valid under the axioms \textbf{C2}, \textbf{C3}, \textbf{C4}, \textbf{SV1}, \textbf{SV2}, \textbf{RW2}, \textbf{RW3}, \textbf{RW5}, \textbf{RW6}, \textbf{RW7} and \textbf{MP}.
\end{lemma}
This means that the proof outline holds for every memory model that satisfies the axioms named. 

To prove this lemma, we have to check every Hoare-triple that we need for local and global correctness (see Def.~\ref{def:correct}). 
Starting with local correctness, Table~\ref{tab:local} gives us an overview of every Hoare-triple we need to prove. 
In there we see which proofs require which rules and thus which axioms. 
For better readability, we have omitted the use of the rules \textsc{Mono, Conj} and \textsc{Disj}. 
One of the Hoare-triples is
\begin{align*}
    {\color{mycolor} \{r_1\neq 1\cup \langle y=1\rangle^{\sf S}[x=1]_3 \}}\ r_2:=^{\sf RS}_3 y\ {\color{mycolor} \{r_1\neq 1\cup r_2\neq 1\cup [x=1]_3\}}
\end{align*}
which we already proved at the end of Section~\ref{sec:axiom}. 
With our novel proof rules at hand we can show its validity with a fewer number of steps. 
As written in Table~\ref{tab:local} we need the rules \textsc{ReadReg} and \textsc{ConRead2}. 
The first one tells us
\begin{align*}
    \{r_1\neq 1 \}\ r_2:=^{\sf RS}_3 y\ \{r_1\neq 1\}
\end{align*}
and with \textsc{ConRead2} we get
\begin{align*}
    \{\langle y=1\rangle^{\sf S}[x=1]_3 \}\ r_2:=^{\sf RS}_3 y\ \{r_2\neq 1\cup [x=1]_3\}
\end{align*}
Applying the \textsc{Disj}-rule we are done. 
Analogously we can now prove every Hoare-triple. 
In this way we need significantly fewer steps to prove one triple than we did in Section~\ref{sec:axiom}. 
Hence the entire correctness proof (which contains the proof of 31 Hoare-triple for Figure~\ref{fig:wrc})
is easier and shorter to prove,
simply by applying the abstract proof rules. An overview of all the rules used for global correctness is given in Table~\ref{tab:global}. Due to the non-interference condition in Owicki-Gries style proofs, there are still a number of proof steps to be done, however significantly fewer than on the level of axioms. 
The number of proof steps could furthermore be reduced by employing a compositional proof technique like rely-guarantee reasoning~\cite{DBLP:journals/fac/XuRH97}. For this, the same proof rules are sound. 

\section{Related Work}
\label{sec:related}

There are a number of approaches which propose novel program logics  for weak memory models. The view-based logic we employ here has first appeared in \cite{DBLP:conf/ppopp/DohertyDWD19,ecoop20} for C11 RAR and has then been generalized to the generic reasoning approach in \cite{TOCL}. 
The work in~\cite{DBLP:conf/esop/BilaDLRW22} uses (and extends) view-based assertions to persistent memory, but does not develop a memory model independent technique. Similarly, Lahav et al.~\cite{CAV23} propose a new program logic for the strong-release-acquire model of 
\cite{DBLP:journals/toplas/LahavB22} and employ rely-guarantee reasoning. While the rely-guarantee framework is independent of a concrete memory model, the program logic is not.   

Besides that there are verification techniques which are applicable to several memory models. Alglave and Cousot \cite{DBLP:conf/popl/AlglaveC17} present an invariance proof method which shows that a given program is correct w.r.t.~a given memory model and an invariant specification of that program. It does so by first proving that a so-called communication specification is sufficient for the program's invariant. 
If a memory model guarantees the communication, the program is correct under that model. 
Ponce de Leon et al.~\cite{DBLP:conf/fmcad/LeonFHM18} and Gavrilenko et al.~\cite{DBLP:conf/cav/GavrilenkoLFHM19} present generic bounded model checkers which translate a given program under a given memory model into an SMT formula. They are generic because their input contains not only the program but also the memory model, formalised in CAT as a set of relations.
Kokologiannakis et al.~\cite{DBLP:conf/pldi/Kokologiannakis19} developed a generic model checker that transforms a given program into an execution graph to check its correctness under a given memory model with an axiomatic semantics. 
Colvin~\cite{DBLP:conf/sefm/Colvin21} proposes a special sequential composition operator which mimics the reordering behaviour of many weak memory models. 
Coughlin et al.~\cite{DBLP:conf/fm/CoughlinWS21,10.1145/3574137} discuss rely-guarantee reasoning for weak memory models in general and introduce a specific new verification condition called reordering-interference-freedom. This technique can be instantiated to memory models with a reordering semantics. 

Our approach discussed here lifts the generic reasoning technique of \cite{TOCL} to the syntactic level, allowing to construct proof outlines operating on the level of program instructions and view-based assertions. Thereby, we avoid low-level reasoning about weakest preconditions while still preserving genericity. 

\section{Conclusion}
\label{sec:con}

In this paper, we have proposed high level proof rules lifting the generic reasoning principle of \cite{TOCL} to a more abstract level. Similar to standard Hoare-logic, our proof rules allow to reason on the syntactic level of assertions, departing from the semantic level of weakest preconditions. 
This significantly simplifies reasoning, and moreover allows us to directly see which axioms have been used in a proof. We have exemplified our proof technique by providing a generic correctness proof for the WRC litmus test. 
By the results of~\cite{TOCL} (showing that SC, TSO and C11 RAR instantiate all axioms), this proof is valid for WRC running on a sequentially consistent as well as the TSO and C11 memory models.  

\bibliographystyle{splncs04}
\bibliography{references}

\appendix 
\section{Proofs}
\label{sec:proofs}
Here you can find the soundness proofs of the theorems in Section~\ref{sec:rul}. We prove them by showing that each of the rules is sound. 
\subsection{Proof of Theorem~\ref{th:general}}
\subsubsection{Proof of \sc{True}.}
For every assertion $P$ and every action $a$ we have 
$$P\subseteq \Sigma = \wlp(T(t,a),\Sigma).$$
This holds because of the non-abortion of $\wlp$.
\subsubsection{Proof of \sc{False}.}
For every assertion $P$ and every action $a$ we have
$$\emptyset\subseteq \wlp(T(t,a),P).$$

\subsubsection{Proof of \sc{Mono}.}
Let $\{P_1\} com_t \{ Q_1 \}$, $P_1\supseteq P_2$ and $Q_1\subseteq Q_2$. 
This means $P_2 \subseteq P_1 \subseteq \wlp(T(t,a),Q_1)$ for the action $a$ in $com_t$. 
With the monotonicity property of $\wlp$ we get $\{P_2\} com_t \{ Q_2\}$.

\subsubsection{Proof of \sc{Conj}.}
Let $\{P_1\} com_t \{ Q_1 \}$ and $\{P_2\} com_t \{ Q_2 \}$. 
For the action $a$ in $com_t$ this means
$P_1 \subseteq \wlp(T(t,a),Q_1) \text{ and } P_2 \subseteq \wlp(T(t,a),Q_2).$ 
Using the conjunctivity of $\wlp$ we get $P_1 \cap P_2 \subseteq \wlp(T(t,a),Q_1 \cap Q_2).$

\subsubsection{Proof of \sc{Disj}.}
Let $\{P_1\} com_t \{ Q_1 \}$ and $\{P_2\} com_t \{ Q_2 \}$. 
For the action $a$ in $com_t$ this means
$P_1 \subseteq \wlp(T(t,a),Q_1) \text{ and } P_2 \subseteq \wlp(T(t,a),Q_2).$ 
Using the disjunctivity of $\wlp$ we get $P_1 \cup P_2 \subseteq \wlp(T(t,a),Q_1 \cup Q_2).$

\subsection{Proof of Theorem \ref{th:fence}}
\subsubsection{Proof of \sc{Fence1}.}
Follows directly from Lemma \ref{lem:fencestable}.

\subsubsection{Proof of \sc{Fence2}.}
Using the FNC axiom and the conjunctivity of $\wlp$ we get 
\begin{align*}
    x_{\uparrow t} \subseteq \wlp(T(t,fence), x_{\uparrow t'}).
\end{align*}

\subsubsection{Proof of \sc{Fence3}.}
Assume the axioms C2, C3, SV1, SV2, RW6 and FNC hold. We devide the proof of 
$$[x=v]_t \subseteq \wlp(T(t,fence), [x=v]_{t'})$$
into the following parts 
\begin{itemize}
    \item[(i)]  $x_{\uparrow t} \subseteq \wlp(T(t,fence), x_{\uparrow t'})$
    \item[(ii)] $[x = v]_t \subseteq \wlp(T(t,fence), [x = v]_t)$
    \item[(iii)]$[x = v]_t \cap x_{\uparrow t'} \subseteq [x = v]_{t'}$
\end{itemize}
(i) follows from the proof of \textsc{Fence 2} (for which we need FNC) and (ii) from the proof of \textsc{Fence 1} (for which we need C3, SV1 and SV2). 
Hence we only prove (iii). 
Because of the axiom RW6 for $\sigma \in [x = v]_t \cap x_{\uparrow t'}$ exists a $u\in \Val$ such that $\sigma \in [x=v]_t \cap x_{\uparrow t'} \cap [x \equiv u]_{t'}$. 
Since C2 also holds, $u=v$ and therefore $\sigma\in [x=v]_t \cap x_{\uparrow t'} \cap [x \equiv v]_{t'}\subseteq [x=v]_{t'}$.
\\\\
With (i), (ii) and the conjunctivity of $\wlp$ we get
$[x = v]_t \subseteq \wlp(T(t,fence), [x = v]_t \cap x_{\uparrow t'})$
and because of (iii) and the monotonicity we have
\begin{align*}
    [x = v]_t \subseteq \wlp(T(t,fence), [x = v]_{t'}).
\end{align*}

\subsection{Proof of Theorem \ref{th:read}}
Let $v'\in\Val$ be the value read. 
\subsubsection{Proof of \sc{Read1}.}
Follows directly from Lemma \ref{lem:readstable}.

\subsubsection{Proof of \sc{Read2}.}
\begin{itemize}
    \item Case $v'\neq v$: We get
    \begin{align*}
        \wlp(T(t,rd(x,r,v')),r\neq v)    &\supseteq  (r\neq v[r:=v'])\\
                                        &=          \Sigma\\
                                        &\supseteq  [x\not\approx v]_t
    \end{align*}
    \item Case $v'=v$: Then because of the monotonicity of $\wlp$
    \begin{align*}
        [x\not\approx v]_t  &=          \wlp(T(t,rd(x,r,v)),\emptyset)\\
                            &\subseteq  \wlp(T(t,rd(x,r,v')), r\neq v')
    \end{align*}
\end{itemize}

\subsubsection{Proof of \sc{Read3}.}
\begin{itemize}
    \item Case $v'=v$: We get 
    \begin{align*}
        \wlp(T(t,rd(x,r,v')),r= v)   &\supseteq  (r=v[r:=v'])\\
                                    &=          \Sigma\\
                                    &\supseteq  [x=v]_t
    \end{align*}
    \item Case $v'\neq v$: Then because of the monotonicity of $\wlp$
    \begin{align*}
        [x=v]_t &\subseteq  [x\not\approx v']_t\\
                &=          \wlp(T(t,rd(x,r,v')),\emptyset)\\
                &\subseteq  \wlp(T(t, rd(x,r,v')), r=v)
    \end{align*}
\end{itemize}

\subsubsection{Proof of \sc{ConRead1}.}
\begin{itemize}
    \item Case $v'\neq v$: We get with the monotonicity of $\wlp$
    \begin{align*}
        \wlp(T(t,rd(x,r,v')),r\neq v \cup [x=v]_t)   &\supseteq  \wlp(T(t,rd(x,r,v')),r\neq v)\\
                                                    &\supseteq  (r\neq v[r:=v'])\\
                                                    &=          \Sigma\\
                                                    &\supseteq  \langle x=v\rangle[x= v]_t
    \end{align*}
    \item Case $v'=v$: Then because of the monotonicity of $\wlp$
    \begin{align*}
        \langle x=v\rangle[x= v]_t   &=          \wlp(T(t,rd(x,r,v)),[x=v]_t)\\
                                        &\subseteq  \wlp(T(t,rd(x,r,v')), r\neq v \cup [x=v]_t)
    \end{align*}
\end{itemize}

\subsubsection{Proof of \sc{ConRead2}.}
\begin{itemize}
    \item Case $v'\neq v$: We get with the monotonicity of $\wlp$
    \begin{align*}
        \wlp(T(t,rd^{\sf RS}(x,r,v')),r\neq v \cup [y=u]_t) &\supseteq  \wlp(T(t,rd^{\sf RS}(x,r,v')),r\neq v)\\
                                                            &\supseteq  (r\neq v[r:=v'])\\
                                                            &=          \Sigma\\
                                                            &\supseteq  \langle x=v\rangle^{\sf S}[y=u]_t
    \end{align*}
    \item Case $v'=v$: Then because of the monotonicity of $\wlp$
    \begin{align*}
        \langle x=v\rangle^{\sf S}[y=u]_t   &=          \wlp(T(t,rd^{\sf RS}(x,r,v)),[y=u]_t)\\
                                    &\subseteq  \wlp(T(t,rd^{\sf RS}(x,r,v')), r\neq v \cup [y=u]_t)
    \end{align*}
\end{itemize}

\subsubsection{Proof of \sc{ReadReg}.}
We have
\begin{align*}
    P = P[r:= v'] \subseteq \wlp(T(t, rd(x,r,v')),P)
\end{align*}

\subsubsection{Proof of \sc{LocRead}.}
Let $P$ be an assertion with $r\not\in reg(P)$. The local program semantics is described in \cite{pso-generic, TOCL}. There we can see that $r:=\e$ will only change the value of $r$. Since $r\not\in reg(P)$ every state in $P$ is still in $P$ after the action.

\subsection{Proof of Theorem \ref{th:write}}
Let $\llbracket E \rrbracket =v$.
\subsubsection{Proof of \sc{Write1}.}
Assume the axioms C3 and SV1 hold. 
Because of the properties of $\wlp$ and $\beta$, we get
\begin{align*}
    [y\not\approx u]_{t'}   &=          \wlp(T(t',rd(y,r,u)),\emptyset)\\
                            &\subseteq  \wlp(T(t',rd(y,r,u))\fcmp\interf(t,wr(x,v)),\emptyset)\\
                            &\subseteq  \wlp(\interf(t,wr(x,v))\fcmp T(t',rd(y,r,u)),\emptyset)\\
                            &=          \wlp(\interf(t,wr(x,v)),\wlp(T(t',rd(y,r,u)),\emptyset))\\
                            &=          \wlp(\interf(t,wr(x,v)),[y\not\approx u]_{t'})\\
                            &\subseteq  \wlp(\beta\fcmp\interf(t,wr(x,v))\fcmp\beta,[y\not\approx u]_{t'})\\
                            &\subseteq  \wlp(T(t,wr(x,v)),[y\not\approx u]_{t'})
\end{align*}

\subsubsection{Proof of \sc{Write2}.}
Assume the axioms C3 and SV1 hold. 
Because of \textsc{Write1} we have for every $u'\neq u$
$\{[y\not\approx u']_{t'}\}x:=_t E\{[y\not\approx u']_{t'}\}$.
With \textsc{Conj} we get 
$\{[y\equiv u]_{t'}\}x:=_t E\{[y\equiv u]_{t'}\}$.

\subsubsection{Proof of \sc{Write3}.}
Assume the axioms C3 and SV2 hold. 
Because of the properties of $\wlp$ and $\beta$, we get
\begin{align*}
    y_{\uparrow t'} \subseteq  \wlp(\interf(t,wr(x,v)),y_{\uparrow t'})
                    \subseteq  \wlp (T(t,wr(x,v)),y_{\uparrow t'})
\end{align*}

\subsubsection{Proof of \sc{Write4}.}
Assume the axioms C3, SV1 and SV2 hold. 
Because of the rules \textsc{Write2} and \textsc{Write3} we have 
$$\{[y\equiv u]_{t'}\}x:=_t E\{[y\equiv u]_{t'}\} \text{ and } \{y_{\uparrow t'}\}x:=_t E\{y_{\uparrow t'}\}$$
Then with \textsc{Conj} we get
$\{[y= u]_{t'}\}x:=_t E\{[y= u]_{t'}\}$.

\subsubsection{Proof of \sc{Write5}.}
Assume the axioms C3 and C4 hold. 
Because of the properties of $\wlp$ and $\beta$, we get
\begin{align*}
    x_{\uparrow t}  \subseteq  \wlp(\interf(t,wr(x,v)),x_{\uparrow t})
                    \subseteq  \wlp (T(t,wr(x,v)),x_{\uparrow t})
\end{align*}

\subsubsection{Proof of \sc{Write6}.}
Assume the axioms C3, C4, RW5 and RW6 hold.
Then we know, because of \textsc{Write5}, RW5 and RW6
\begin{itemize}
    \item $x_{\uparrow t} \subseteq  \wlp (T(t,wr(x,v)),x_{\uparrow t})$
    \item $\Sigma\subseteq \wlp (T(t, wr(x,v)), \dom(T(t,\Rd_{|x}[v])))$
    \item $x_{\uparrow t} \subseteq \bigcup_{u'\in\Val}[x\equiv u']_t$
\end{itemize}
Using the conjunctivity of $\wlp$, we get 
\begin{align*}
    [x=u]_t &\subseteq  \wlp(T(t, wr(x,v)),x_{\uparrow t}\cap \dom(T(t,\Rd_{|x}[v])) \cap \bigcup_{u'\in\Val}[x\equiv u']_t)\\
            &=          \wlp(T(t, wr(x,v)),x_{\uparrow t}\cap [x\equiv v]_t)\\
            &=          \wlp(T(t, wr(x,v)),[x=v]_t)
\end{align*}

\subsubsection{Proof of \sc{ConWrite1}.}
Assume the axioms C2, C3, C4, SV1, SV2, RW2, RW3, RW5, RW6 and RW7 hold.
RW7 says that $$x_{\uparrow t} \cap [x\not\approx v]_{t'} \subseteq \wlp(T(t,wr(x,v)),\wlp(T(t',rd(x,r,v)),x_{\uparrow t'}))$$
Additionally we know, because of \textsc{Write6} and \textsc{Read1}
\begin{align*}
    [x=u]_t     &\subseteq  \wlp(T(t,wr(x,v)),[x=v]_t)\\
                &\subseteq  \wlp(T(t,wr(x,v)),\wlp(T(t',rd(x,r,v))[x=v]_t))
\end{align*}
With C2 and RW6 we get $[x=v]_t\cap x_{\uparrow t'} \subseteq [x=v]_{t'}$ and therefore because of $\wlp$'s properties
\begin{align*}
    [x=u]_t\cap [x\not\approx v]_{t'}   &\subseteq  \wlp(T(t,wr(x,v)),\wlp(T(t',rd(x,r,v))[x=v]_t\cap x_{\uparrow t'}))\\
                                        &\subseteq  \wlp(T(t,wr(x,v)),\wlp(T(t',rd(x,r,v))[x=v]_{t'}))\\
                                        &\subseteq  \wlp(T(t,wr(x,v)),\langle x=v\rangle[x=v]_t)
\end{align*}

\subsubsection{Proof of \sc{ConWrite2}.} 
Assume the axioms C2, C3, SV1, SV2, RW2, RW3, RW6 and MP hold. To show that 
$$[x\not\approx v]_{t'} \cap [y=u]_t \subseteq \wlp(T(t,wr^{\sf WS}(x,v)),\langle x=v \rangle^{\sf S} [y=u]_{t'})$$
holds for $x\neq y$ and $t\neq t'$, we first prove
\begin{itemize}
    \item[(i)]  $y_{\uparrow t} \cap [x\not\approx v]_{t'} \subseteq \wlp(T(t,wr^{\sf WS}(x,v)),\wlp(T(t',rd^{\sf RS}(x,r,v)), y_{\uparrow t'}))$
    \item[(ii)] $y_{\uparrow t} \subseteq \wlp(T(t,wr^{\sf WS}(x,v)),\wlp(T(t',rd^{\sf RS}(x,r,v)), y_{\uparrow t}))$
    \item[(iii)]$[y\equiv u]_t \subseteq \wlp(T(t,wr^{\sf WS}(x,v)),\wlp(T(t',rd^{\sf RS}(x,r,v)), [y\equiv u]_t))$
    \item[(iv)] $y_{\uparrow t} \cap [y \equiv u]_t \cap y_{\uparrow t'} \subseteq [y = u]_{t'}$
    \item [(v)] $[y=u]_{t'}\subseteq \langle x=v \rangle^{\sf S}[y=u]_{t'}$
\end{itemize}
Then because of the conjunctivity of $\wlp$ and (i)-(iii) we get 
\begin{align*}
    & [x\not\approx v]_{t'} \cap [y=u]_t \\
    & \subseteq \wlp(T(t,wr^{\sf WS}(x,v)),\wlp(T(t',rd^{\sf RS}(x,r,v)), y_{\uparrow t} \cap [y \equiv u]_t \cap y_{\uparrow t'}))
\end{align*}
and, because of the monotonicity, (iv) and (v), the rule is proven. 
Therefore we now prove (i)-(v).
\begin{itemize}
    \item[(i)] With the MP Axion and the conjunctivity and monotonicity of $\wlp$ we get 
    \begin{align*}
        & \wlp(T(t,wr^{\sf WS}(x,v)),\wlp(T(t',rd^{\sf RS}(x,r,v)), y_{\uparrow t'})) \\
        & \supseteq y_{\uparrow t} \cap \wlp(T(t',rd^{\sf RS}(x,r,v)),y_{\uparrow t'}) \\
        & \supseteq y_{\uparrow t} \cap \wlp(T(t',rd^{\sf RS}(x,r,v)),\emptyset) \\
        & \supseteq y_{\uparrow t} \cap [x\not\approx v]_{t'}
    \end{align*}
    \item[(ii)] Using the proofs of \textsc{Read 1} (for which we need C3, SV1, Sv2, RW2 and RW3) and \textsc{Write 3} (for which we need C3 and SV2) we get 
    \begin{align*}
        & \wlp(T(t,wr^{\sf WS}(x,v)),\wlp(T(t',rd^{\sf RS}(x,r,v)), y_{\uparrow t}))\\
        & \supseteq \wlp(T(t', wr^{\sf WS}(x,v)), y_{\uparrow t})\\
        & \supseteq y_{\uparrow t}
    \end{align*}
    \item[(iii)]Using the proofs of \textsc{Read 1} (for which we need C3, SV1, SV2, RW2 and RW3) and \textsc{Write 2} (for which we need C3 and SV1) we get 
    \begin{align*}
        & \wlp(T(t,wr^{\sf WS}(x,v)),\wlp(T(t',rd^{\sf RS}(x,r,v)), [y\equiv u]_t))\\
        & \supseteq \wlp(T(t,wr^{\sf WS}(x,v)),[y\equiv u]_t)\\
        & \supseteq [y\equiv u]_t
    \end{align*}
    \item[(iv)] Because of the axiom RW6 for $\sigma \in y_{\uparrow t} \cap [y \equiv u]_t \cap y_{\uparrow t'}$ exists a $u'\in \Val$ such that $\sigma \in y_{\uparrow t} \cap [y \equiv u]_t \cap y_{\uparrow t'} \cap [y \equiv u']_{t'}$. Since C2 also holds, $u=u'$ and therefore $\sigma\in y_{\uparrow t} \cap [y \equiv u]_t \cap y_{\uparrow t'} \cap [y \equiv u]_{t'}\subseteq [y=u]_{t'}$
    \item[(v)]  Because of the proof of \textsc{Read 1} we get
    \begin{align*}
        & [y=u]_{t'} \\
        & \subseteq \wlp(T(t', rd^{\sf RS}(x,r,v)), [y=u]_{t'}) \\
        & = \langle x=v \rangle^{\sf S}[y=u]_{t'}
    \end{align*}
\end{itemize}

\subsubsection{Proof of \sc{WriteReg}.}
In \cite{pso-generic,TOCL} the local program semantics (which is independent of the given memory model) is described. There we can see that a write action will not effect the local state. Hence the rule holds.

\end{document}